# Modification of streaming potential by precipitation of calcite in a sand–water system: laboratory measurements in the pH range from 4 to 12

Xavier Guichet,[1] Laurence Jouniaux[2] and Nicole Catel[3]

[1]*Laboratoire des mécanismes de transfert en géologie, UMR 5563, 38 rue des 36 ponts, 31400 Toulouse, France.*
E-mail: Jouniaux@eost.u-strasbg.fr
[2]*Institut de Physique du Globe de Strasbourg, UMR 7516, 5 rue René Descartes, 67000 Strasbourg, France*
[3]*École Normale Supérieure, Laboratoire de Géologie, UMR 8538, 24 Rue Lhomond, Paris, France*

**SUMMARY**

Spontaneous potentials associated with volcanic activity are often interpreted by means of the electrokinetic potential, which is usually positive in the flow direction (i.e. Zeta potential of the rock is negative). The water–rock interactions in hydrothermal zones alter the primary minerals leading to the formation of secondary minerals. This work addresses the study of calcite precipitation in a sand composed of 98 per cent quartz and 2 per cent calcite using streaming potential measurements. The precipitation of calcite as a secondary mineral phase, inferred by high calcite saturation indices and by a fall in permeability, has a significant effect on the electrokinetic behaviour, leading to a significant reduction in the Zeta potential (in absolute value) and even a change in sign.

The measured decrease in Zeta potential from $-16$ mV to $-27 \pm 4$ mV takes place as the pH rises from 4 to 7, while it remains constant at $-25 \pm 1$ mV as the pH increases from 8 to 10.5. For pH higher than 10.5, calcite precipitates and is expected to coat the quartz surface. The measured Zeta potential vary from $-17$ to $+8$ mV for pH ranging from 10.6 to 11.7 depending on the amount of precipitated calcite indicated by the decrease in permeability.

The observed change in sign of the electrical surface potential rules out the usual qualitative interpretation of SP anomalies in order to determine fluid circulations, even at pH lower than 9 if calcite is widely present as a secondary mineral phase, since the electrical surface potential of calcite depends also on $CO_2$ partial pressure and $[Ca^{2+}]$. Therefore, SP anomalies as measured in hydrothermal field, without mineralogical analyses of hydrothermal deposits, and without geochemical fluid survey, should be interpreted with caution.

## 1 INTRODUCTION

Spontaneous potential (SP) is an electrical geophysical method that measures naturally occurring voltage fields on the Earth's surface. Measurements of SP in the field are performed in various different geophysical contexts and are often interpreted by means of the electrokinetic or streaming potential, which represents the electrical potential induced by fluid flow through rock. In tectonically active zones, SP anomalies may be explained by a mechanism involving fluid movements between reservoirs triggered by strain perturbations (Bernard 1992), or by changes in pore pressure in fault zones (Fenoglio *et al.* 1995). The negative anomalies observed on the flanks of volcanoes are interpreted in terms of downward percolating rainfall, and can be linked to the depth of the water table (Zablocki 1978; Aubert *et al.* 1993). The positive anomalies observed in active volcanic areas originate from upward convective flows (Zablocki 1978; Aubert & Kieffer 1984), and are used to define hydrothermal zones (Corwin & Hoover 1979; Ishido *et al.* 1997; Michel & Zlotnicki 1998; Lénat *et al.* 2000; Finizola *et al.* 2002, 2003, 2004; Hase *et al.* 2005). Recent experiments have demonstrated that time variations of SP can be unambiguously identified and associated with time-varying fluid flow in geophysical systems from metric to kilometric scales (Perrier *et al.* 1998; Doussan *et al.* 2002; Pinettes *et al.* 2002; Trique *et al.* 2002), thanks to the improvement in long-term monitoring SP (Perrier *et al.* 1997; Perrier & Pant 2005). Electroseismic surveys are also based on electrokinetic phenomena and can be used to detect small changes in rock properties at interfaces, such as permeability and saturation state (Bordes *et al.* 2006; Garambois & Dietrich 2001, 2002; Beamish 1999). Moreover, the borehole electrokinetic response due to water injection (Marquis *et al.* 2002; Darnet *et al.* 2004) can yield an estimate of the fracture aperture (Hunt & Worthington 2000) or permeability (Murakami *et al.* 2001); and the inversion of SP anomalies (Gibert & Pessel 2001; Sailhac & Marquis 2001; Sailhac *et al.* 2004) can yield an estimate of aquifer hydraulic properties (Darnet *et al.* 2003). Measurements of SP have also been used in karst areas in the detection of groundwater flows (Erchul & Slifer 1987; Wanfang *et al.* 1999). Modelling of all these observations requires a good understanding of electrokinetic phenomena. We emphasize that it is fundamental to check for the presence of calcite in order to interpret field measurements of SP, such as in the case of streaming potentials in underground limestone quarries (Morat *et al.* 1995) or for the preservation of historical monuments (Pisarenko *et al.* 1996). Indeed the presence of calcite can induce a change in the streaming potential sign, which is generally positive in the flow direction (Ishido & Mizutani 1981; Jouniaux & Pozzi 1995a, 1997; Guichet *et al.* 2003; Maineult *et al.* 2004).

In hydrothermal areas, interaction between hot waters and the rocks through which they migrate alters the primary minerals and leads to the formation of secondary minerals. These processes result in changes in physical and chemical properties of the system. Karst areas are generally underlain by soluble calcareous rocks. Sinkholes, caves, result from dissolution of primary rocks. Sealing of fractured rocks can also occur, resulting in the precipitation of secondary minerals. Davis & Kent (1990) reported that the surface chemical properties of natural materials are modified by secondary minerals that are usually present as a minor fraction of the whole sample. We decided to focus our attention on calcite as a secondary mineral, because (i) Calcite is a common mineral constituent of limestones, and is a secondary mineral in numerous geological contexts, and (ii) the electrical properties of the calcite–water interface is still a controversial topic.

(i) In natural geophysical systems, calcite is a common mineral phase, modifying ground- and surface-water compositions. It acts as a buffer for pH in ground and surface waters, being able to modify the water chemistry because of its low solubility and rapid precipitation–dissolution kinetics (Langmuir 1971; Sigg *et al.* 2000). Waters in or close to equilibrium with calcite contain large amounts of $Ca^{2+}$ and $HCO_3^-$; to a first approximation the concentration of $HCO_3^-$ is twice the concentration of $Ca^{2+}$ (Sigg *et al.* 2000). Many natural rivers show a calcite oversaturation with a $CO_2$ partial pressure ranging from atmospheric $CO_2$ partial pressure to $10^{-2}$ atm (Fig. 1). Thus, calcite acts as a buffer for many natural waters, and is expected to precipitate in many natural systems.

Calcite is present as a secondary mineral phase in volcanic terrains; Deutsch *et al.* (1982) showed that the water properties of the Columbia plateau basalt aquifer are determined by secondary calcite and Robert (2001) observed calcite in basalts from Northern Ireland, while Stoffers & Botz (1994) showed calcite formation associated with hydrothermal fluids emerging from the floor of Lake Tanganyika. In general the precipitation of secondary phases during the alteration of volcanic rocks is related to the thermal behaviour of the hydrothermal system. With decreasing temperature, the following crystallization sequence is observed: smectite → zeolite → calcite. These observations are based on field studies (Westercamp 1981; Robert *et al.* 1988) and also on experimental results (Robert & Goffé 1993). (Henley & Ellis 1983) gave a summary of the temperature range over which calcite alteration mineral has been observed in high-temperature hydrothermal systems associated with volcanism; calcite alteration mineral occurs from 100°C up to 300°C.

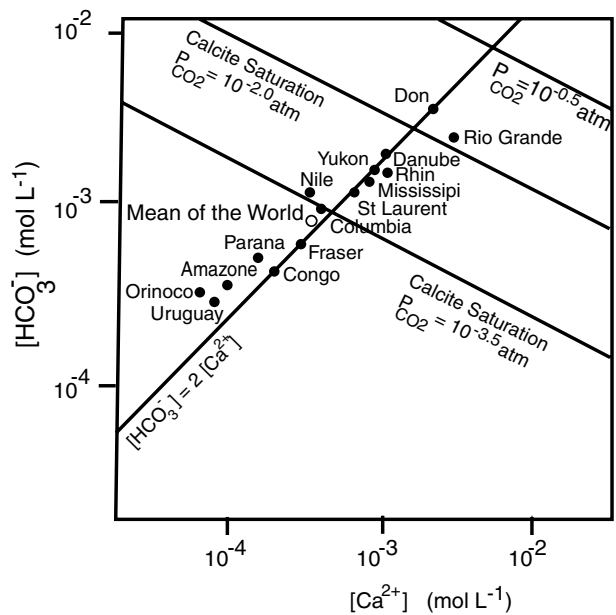

**Figure 1.** The concentration of $HCO_3^-$ versus the concentration of $Ca^{2+}$ for the main rivers of the world. The bold line shows the electroneutrality of waters buffered by calcite, for which $HCO_3^-$ and $Ca^{2+}$ are the main chemical species. The thin lines show calcite saturation for several $CO_2$ partial pressures (1 atm = 1.013 Pa). The waters of the main rivers fall into the bold line. (from Sigg *et al.* 2000).

(ii) Geochemists have investigated sorption processes onto calcite surfaces (Douglas & Walker 1950; Foxall *et al.* 1979; Zachara *et al.* 1991; Cicerone *et al.* 1992), as well as interactions at the molecular scale (Stipp & Hochella 1991). Nevertheless, the electrical properties of the calcite–water interface is still a controversial topic (Fig. 2), and the calcite surface reactions appear to be more complex than those involving silicates. Electrokinetic measurements are sometimes contradictory, yielding different results on natural calcite and on synthetic calcite, as well as depending on $CO_2$ partial pressure. For example in natural systems, the $Ca^{2+}$ concentration ranges from $10^{-2}$ to $10^{-4}$ mol l$^{-1}$ for systems where calcite acts as a buffer (Sigg *et al.* 2000). Measured calcite $\zeta$ potential ranges from 17 mV to $-11$ mV, while $Ca^{2+}$ concentration ranges from $10^{-2}$ to $10^{-3.3}$ mol L$^{-1}$ (Cicerone *et al.* 1992).

The two last decades, laboratory measurements have been performed to understand the variations of the streaming potential in fluid chemistry for various silicate minerals (Ishido & Mizutani 1981; Jouniaux & Pozzi 1995b; Lorne *et al.* 1999a,b; Pengra *et al.* 1999; Jouniaux *et al.* 2000; Guichet & Zuddas 2003; Guichet *et al.* 2003; Perrier & Froidefond 2003; Maineult *et al.* 2005; Lorne *et al.* 1999a) also reported the streaming potentials of four carbonate rocks at pH 8. However, it is surprising that the effect of secondary minerals in electrokinetic properties has not been studied. Modelling of SP observations in either hydrothermal or karst areas requires a good understanding of electrokinetic phenomena in dynamic systems where dissolution/precipitation can occur. In this study, we report streaming potential measurements on sand and show that precipitation of calcite as a secondary mineral phase has a significant effect on the electrokinetic behaviour: that is, the $\zeta$-potential is reduced significantly (in absolute value) and can even change sign. Finally our measured surface potentials are compared to calculated surface potentials using a triple-layer model (TLM).

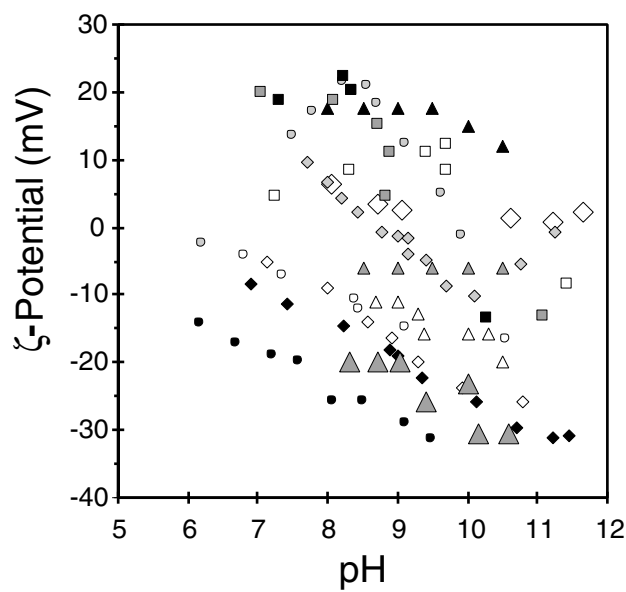

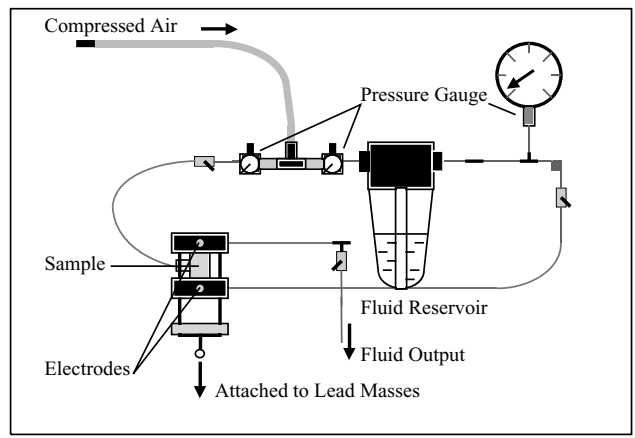

**Figure 3.** The water pressure is controlled by applying compressed air to a water reservoir at one end of the sample, while the other end is maintained at atmospheric pressure. The electric potential difference is measured between the ends of the sample by two silver-chloride non-polarizable electrodes: water saturated porous ceramics are used to prevent any gas or air bubble from being in contact with the electrode and remain saturated to keep the electrical contact. The electrodes are placed in the water circuit near the ends of the sample but not in the circulating water flow to avoid electrical noise due to water movement near the electrodes (Ahmad 1964). When calcite is precipitated during the experiments, the electrical potential difference measured by the electrodes without any fluid flow remains small (1–4 mV). The two electrodes are connected by coaxial wires to a high-input impedance voltmeter (>10 GΩ). The input impedance of the voltmeter is several orders of magnitude greater than the resistance of the sample (e.g. it is 16 kΩ when water conductivity is 0.036 S m$^{-1}$), thus allowing accurate measurement of the electric potential.

**Figure 2.** Variation of the $\zeta$-potential of calcite as a function of pH. Squares show measurements of $\zeta$-potential of Iceland Spar (empty squares after a few minutes, grey squares after one week, black squares after two months) performed by (Somasundaran & Agar 1967); Circles show measurements of $\zeta$-potential performed by (Vdovic 2001) with natural samples (empty circles: limestone, black circles: lake sediment) and with synthetic sample (grey circles). Diamonds show measurements of $\zeta$-potential of synthetic calcite performed by (Thompson & Pownall 1989) with various electrolytes (Black diamonds: NaCl ($5 \times 10^{-3}$ mol L$^{-1}$), Empty diamonds: NaCl ($5 \times 10^{-3}$ mol $L^{-1}$)/NaHCO$_3$ ($10^{-3}$ mol L$^{-1}$), Big Empty diamonds CaCl$_2$ ($5 \times 10^{-4}$ mol L$^{-1}$), pH values adjusted by additions of HCl/NaOH; Grey diamonds: NaCl ($5 \times 10^{-3}$ mol L$^{-1}$)/NaHCO$_3$ ($10^{-3}$ mol L$^{-1}$), pH values adjusted by additions of Ca(OH)$_2$. Triangles show measurements of $\zeta$-potential performed by (Cicerone *et al.* 1992) with a natural sample of Ficopomatus Enigmaticus (Big grey triangles) and with synthetic sample for various electrolytes (KCl ($10^{-3}$ mol L$^{-1}$) solutions containing different amounts of added CaCl$_2$ (0 mol L$^{-1}$, empty triangles; $10^{-3}$ mol L$^{-1}$ Grey triangles; $10^{-2}$ mol L$^{-1}$ Black triangles).

## 2 SAMPLE AND METHODS

### 2.1 Electrical measurements

We perform laboratory streaming potential measurements of a sand composed of 98 per cent quartz and 2 per cent calcite using CaCl$_2$ solutions made up in distilled water for pH in the range 4–12. The sand is sieved, yielding a grain size ranging from $100 \times 10^{-6}$ m to $500 \times 10^{-6}$ m. Experiments are carried out at room temperature, which varies from 14.9°C to 19.5°C. A solution of CaCl$_2$, $5 \times 10^{-4}$ mol l$^{-1}$ is then prepared. The water used for making all solutions is freshly distilled with a MilliQ system. The solutions are made up and stored in glassware, thus atmospheric CO$_2$ does not equilibrate with the solution, which remains undersaturated with respect to atmospheric CO$_2$. Electrical surface potential of calcite can range from positive to negative values depending on CO$_2$ partial pressure, pH, and [Ca$^{2+}$] (Pokrovsky *et al.* 1999). At the atmospheric CO$_2$ partial pressure a pH corresponding to a zero surface charge can be observed. Since we do not observe any point of zero surface charge, the CO$_2$ partial pressure of our experiments must be lower than the atmospheric CO$_2$ partial pressure (Van Cappellen *et al.* 1993). The pH of the solution is adjusted with Ca(OH)$_2$ in the range 7–11.7, and with HCl in the range 4.4–7.

The experiment setup is shown in Fig. 3. The reader can find a detailed description of the apparatus in (Jouniaux *et al.* 2000). In a typical experiment, a 1L-aliquot of aqueous solution from the reservoir is passed through the sample. The output solution is recirculated twice. Reliable measurements involve equilibrium between sand and aqueous solution: the electrical conductivity and the pH of the output solution are measured after the different steps of fluid circulation until the values become constant. Equilibrium is reached after waiting 24 hr (see Appendix A for details). Before measuring the electrokinetic coupling coefficient, the 1L aliquot of aqueous solution is again recirculated twice. A fraction of the output solution is sampled for chemical analysis.

The streaming potential $\Delta V$ is measured when the fluid is forced through the sample by applying a fluid pressure difference $\Delta P$. To check the linearity between the applied pressure difference $\Delta P$ at the sample ends and the measured streaming potential difference $\Delta V$, the solution is made to flow systematically with 5 pressure steps from 0.1 to $0.5 \times 10^5$ Pa. The coupling coefficient $\Delta V / \Delta P$ is determined from the slope of $\Delta V$ against $\Delta P$ (see examples in Fig. 4). The streaming measurements are described in Section 3 and the analysis of the effect of the precipitation of calcite as a secondary mineral phase is made in Section 4.

The permeability of the sample is also derived using Darcy's law. Since the sample permeability is high, more than $10^{-12}$ m$^2$ (see Table 1), the aqueous solution filling the pore space can be easily replaced through fluid circulation by fresh aqueous solution having a new pH. Experiments are performed on three samples of the same sand. The first sample runs are performed in the pH range from 7 to 11.7 adding Ca(OH)$_2$; the second sample runs are performed in the pH range 9–11.7 to have a better analysis in this range. And the third sample runs are performed in the pH range 7–4 adding HCl (Table 1). We do not extend these measurements to pH below 4.4 and above 11.7, because then the ionic strength becomes significantly higher, and the interpretation of the experiments assuming a zeta potential non-dependent on the electrical conductivity is no more valid. Any decrease in permeability is thought to be related to calcite precipitation as discussed in Section 4.

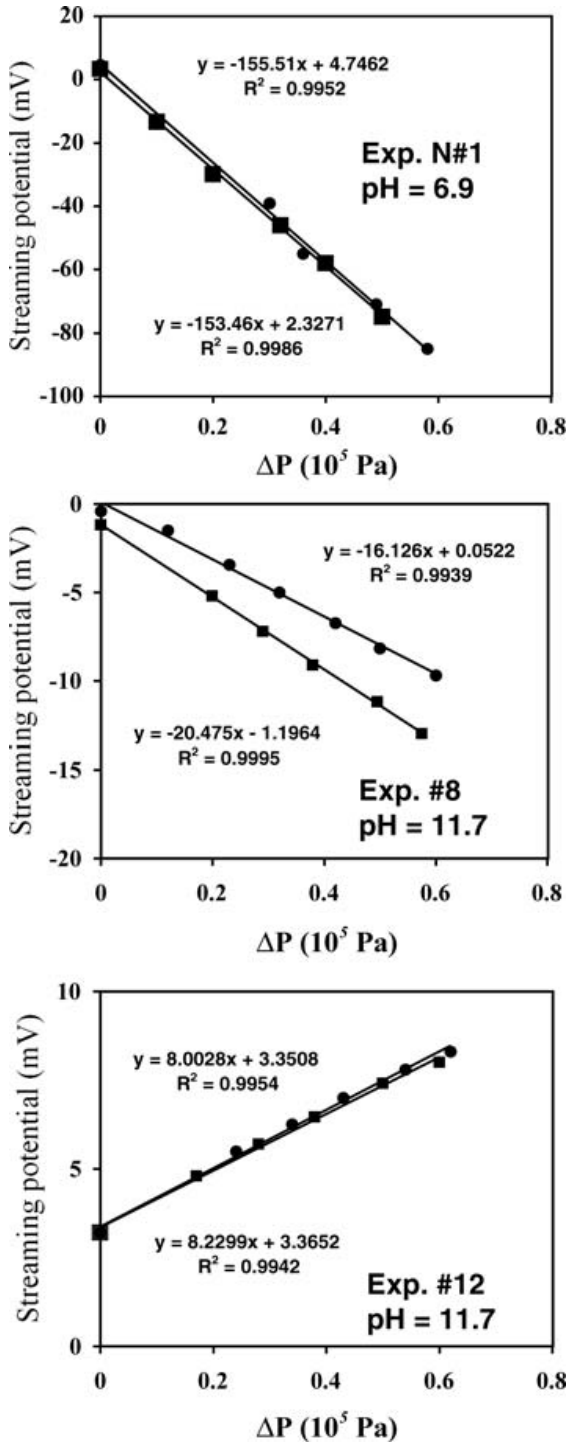

**Figure 4.** Streaming potential measured as a function of applied pressure during experiments 1, 8 and 12.

Furthermore, the electrical conductivity of the sand $\sigma_r$ is measured as a function of the electrical conductivity $\sigma_f$ of the electrolyte in order to obtain the formation factor $F$ ($F = \sigma_f/\sigma_r$) and the surface electrical conductivity (see for instance: Waxman & Smits 1968).

### 2.2 Chemical analyses

It has been emphasized (Pokrovsky & Schott 1999) that there are three potential-controlling parameters of the interfacial properties of carbonates (pH, Alkalinity, [Ca$^{2+}$]), and that in previous carbonates electrokinetic studies, these parameters were unfortunately not measured. Therefore, pH, alkalinity and [Ca$^{2+}$] were measured on the outflow solution collected after each experiment. We also measured Na$^+$ and Cl$^-$. A Perkin-Elmer 2380 spectrometer is used to determine the concentrations of calcium and sodium. Calcium is determined by atomic absorption, and sodium by atomic emission. A Hitachi U 1100 spectrometer is used to determine the concentrations of chlorine and silicon. Chlorine is determined by stable complexation of mercury (II) thiocyanate. Silicon is determined by complexation of molybdic silicide, while alkalinity is determined by colorimetric titration with bromocresol green and methyl red.

### 2.3 $\zeta$-potential calculations

Several models link the electrokinetic coupling coefficient $C_s = \Delta V/\Delta P$ to the physical properties of the fluid, the physical properties of the porous medium and the electric surface potential. We use the expression obtained with equation (11) from (Bussian 1983) and equation (46) from (Revil et al. 1999b):

$$C_s = \frac{\varepsilon\zeta/\eta\sigma_f}{1 + m(F-1)\xi}, \qquad (1)$$

where

(i) $\zeta$ is defined as the electric potential on the shear plane (the closest plane to the rock surface over which fluid is moving)

(ii) $\varepsilon = \varepsilon_{\text{water}}\,\varepsilon_0$ is the electric permittivity of the fluid with $\varepsilon_{\text{water}}$ the relative dielectric constant of the fluid, which is calculated as a function of the temperature according to (Malmberg & Maryott 1956)'s law, and $\varepsilon_0 = 8.84 \times 10^{-12}$ F/m the electric permittivity of vacuum, and $\eta$ is the dynamic shear viscosity of the circulating fluid, which is temperature-dependent, equal to $1.01 \times 10^{-3}$ Pa s at 19.5°C and to $1.14 \times 10^{-3}$ Pa s at 14.9°C.

(iii) $\sigma_f$ is the electrical conductivity of the fluid, and $\xi = \sigma_s/\sigma_f$ with $\sigma_s$ being the surface electrical conductivity of the sample.

(iv) $F$ is the formation factor defined by Archie's first law:

$$F \equiv \Phi^{-m}$$

where $\Phi$ is the porosity, and $m$ the first Archie's exponent (or the cementation exponent) of the rock.

Eq. (1) is used to derive the $\zeta$-potential from our measurements of coupling coefficient $C_s$. Eq. (1) is valid in the high-salinity domain, that is, $\sigma_f \geq 5(m-1)\sigma_s$ (Bussian 1983), yielding a value equal to the high-salinity limit in the model of Revil et al. (1999b) and is also equivalent to the expression given by Jouniaux & Pozzi (1995a) in the case of m being equal to 2.

### 3 RESULTS

We describe here the results, but the interpretation of the electrokinetic behaviour as calcite is precipitated is discussed in Section 4

**Table 1.** Description of the experiments ; temperature, electrical conductivity, and pH of the electrolyte; measured electrokinetic coupling coefficient; $\zeta$-potential inferred from eq. (1) using $F = 4.8$, $\sigma_s = 10^{-4}$ S m$^{-1}$ and $m = 1.3$. Three samples have been used: sample 1 with pH solutions 7 to 11.7; sample 2 with pH solutions from 9 to 11.7; sample 3 with pH solutions from 7 to 4.4.

| | | Aqueous solution properties | | | | Sample properties | |
|---|---|---|---|---|---|---|---|
| Experiment | Sample | Temperature °C | Electrical conductivity mS m$^{-1}$ | pH | Permeability $10^{-12}$ m$^2$ | Coupling Coefficient mV/MPa | $\zeta$-Potential mV |
| 1  | 1 | 19.5 | 12.7 | 6.9  | –           | $-1545 \pm 1$  | $-29 \pm 0.5$ |
| 2  | 1 | 18.5 | 13.3 | 7.0  | $3.9 \pm 0.5$ | $-1173 \pm 11$ | $-23 \pm 1.2$ |
| 3  | 1 | 17.9 | 15.4 | 8.2  | –           | $-1136 \pm 1$  | $-25 \pm 1.4$ |
| 4  | 1 | 17.1 | 15.4 | 8.4  | –           | $-1157 \pm 1$  | $-26 \pm 2$ |
| 5  | 1 | 14.9 | 16.2 | 9.5  | –           | $-1043 \pm 1$  | $-24 \pm 3.3$ |
| 6  | 1 | 15.7 | 17.2 | 10.1 | $4.0 \pm 0.1$ | $-1016 \pm 15$ | $-25 \pm 3.3$ |
| 7  | 1 | 18.8 | 19.0 | 10.7 | $1.5 \pm 0.1$ | $-412 \pm 8$   | $-11 \pm 0.6$ |
| 8  | 1 | 19.5 | 63.3 | 11.7 | $1.7 \pm 0.4$ | $-183 \pm 22$  | $-16 \pm 2.1$ |
| 9  | 2 | 18.3 | 12.1 | 9.0  | $4.4 \pm 0.1$ | $-1348 \pm 8$  | $-24 \pm 1.3$ |
| 10 | 2 | 18.3 | 17.4 | 10.5 | $4.4 \pm 0.1$ | $-1047 \pm 4$  | $-26 \pm 1.3$ |
| 11 | 2 | 18.4 | 26.7 | 11.0 | $3.2 \pm 0.1$ | $-435 \pm 4$   | $-17 \pm 0.9$ |
| 12 | 2 | 19.3 | 72.5 | 11.7 | $0.2 \pm 0.1$ | $81 \pm 1$     | $8 \pm 0.3$ |
| 13 | 3 | 17.0 | 13.2 | 7.5  | $5.6 \pm 0.1$ | $-1587 \pm 12$ | $-30 \pm 2.7$ |
| 14 | 3 | 16.7 | 14.2 | 7.4  | $5.5 \pm 0.2$ | $-1396 \pm 43$ | $-28 \pm 3.3$ |
| 15 | 3 | 16.4 | 15.4 | 7.1  | $5.6 \pm 0.1$ | $-1263 \pm 5$  | $-28 \pm 2.8$ |
| 16 | 3 | 16.4 | 16.9 | 6.6  | $5.5 \pm 0.1$ | $-1055 \pm 3$  | $-26 \pm 2.6$ |
| 17 | 3 | 18.2 | 23.8 | 6.5  | $5.7 \pm 0.1$ | $-650 \pm 1$   | $-22 \pm 1.1$ |
| 18 | 3 | 17.5 | 36.9 | 4.4  | $5.5 \pm 0.1$ | $-304 \pm 16$  | $-16 \pm 1.9$ |
| 19 | 3 | 18.2 | 36.2 | 5.8  | $5.5 \pm 0.1$ | $-314 \pm 2$   | $-16 \pm 0.8$ |
| 20 | 3 | 18.1 | 13.5 | 5.9  | $5.0 \pm 0.1$ | $-1078 \pm 32$ | $-21 \pm 1.7$ |
| 21 | 3 | 15.4 | 12.8 | 6.5  | $4.8 \pm 0.1$ | $-1702 \pm 1$  | $-31 \pm 3.9$ |

–Not measured.

since this interpretation is discussed using the permeability measurements (Section 3.1), the chemical analysis (Section 3.2) and the electrical measurements (Section 3.1).

### 3.1 Electrical properties and permeability of the sand

In order to determine the formation factor, the electrical conductivity of the sand is measured as a function of water electrical conductivity in the range 0.04–0.8 S m$^{-1}$. Electrical measurements yield a formation factor, $F$, of 4.8. The conductivity of the electrolyte after flowing through the sample is never lower than $1.2 \times 10^{-2}$ S m$^{-1}$, which prevents us from determining the electrical surface conductivity of the sand. Revil & Glover (1998) reported a surface conductance of $8 \times 10^{-9}$ S for quartz, independent on the electrolyte conductivity when this latter exceeds $10^{-2}$ S m$^{-1}$. Since the electrical conductivity of the water circulating in the sand is greater than $10^{-2}$ S m$^{-1}$ (see Table 1), we assume that the specific surface conductance of the sand is $8 \times 10^{-9}$ S. This is equivalent to a surface conductivity (i.e. 2× surface conductance divided by grain radius) of about $1.06 \times 10^{-4}$ S m$^{-1}$ for a mean grain diameter of 300 $\mu$m, in agreement with experimental values (Ruffet *et al.* 1991; Lorne *et al.* 1999a). The values of formation factor and surface conductivity given here are used in eq. (1) to obtain $\zeta$-potentials from the electrokinetic measurements.

Fig. 5(a) shows the variation of sample permeability as a function of solution pH. Permeability of the samples is insensitive to pH over the range from 4 to 9, being $4.15 \pm 0.15 \times 10^{-12}$ m$^2$ for sample 1 and $4.37 \pm 0.04 \times 10^{-12}$ m$^2$ for sample 2. The permeability of sample 3 remains constant at $5.4 \pm 0.2 \times 10^{-12}$ m$^2$ for a pH range of 4.4–7.7. The permeability decreases when pH is higher than 9. We observe that the permeability of sample 1 decreases up to $1 \times 10^{-12}$ m$^2$ at pH 11.7, and that the permeability of sample 2 falls to $0.2 \times 10^{-12}$ m$^2$ at pH 11.7. This decrease of permeability can be shown to be related to calcite precipitation.

Fig. 5(b) shows the variation of the aqueous solution electrical conductivity as a function of the solution pH. The water electrical conductivity measurements are shown because the $\zeta$-potential values are interpreted assuming a constant water conductivity. The solution electrical conductivity is constant at about $1.55 \pm 0.20 \times 10^{-2}$ S m$^{-1}$ for pH in the range 5.9–10.7. Under either highly acidic or basic conditions, the solution electrical conductivity increases up to $7 \times 10^{-2}$ S m$^{-1}$ due to the addition of HCl at low pH and Ca(OH)$_2$ at high pH. The variation of the solution electrical conductivity is linked to the variation in the ionic strength (Fig. 5d) as calculated from the results of the chemical analyses (Table 2) and the WATEQ program (Plummer *et al.* 1976). The ionic strength is constant ($I = 2 \pm 0.2 \times 10^{-3}$ mol l$^{-1}$), while pH ranges from 5.5 to 11 except for experiment 17 (Fig. 5d). The ionic strength of experiments 8 and 12 (the most basic experimental conditions, pH 11.7) is not given because the electrical charge is unbalanced in this case.

Fig. 5(c) shows the variation of the $\zeta$ potential derived from eq. (1) for pH in the range 4–12. To calculate the $\zeta$ potential, we use $F = 4.8, \sigma_s = 1.06 \times 10^{-4}$ S m$^{-1}$ and $m = 1.3$, the latter being a typical value for sand (Archie 1942). The formation factor is assumed to be constant, although it could be increased when permeability is decreased, specially for experiment n°12. Since the surface conductivity is small compared to the fluid conductivity, the value of $m(F - 1)\xi$ in the eq. (1) is small compared to 1, so that the $\zeta$-potential derived from eq. (1) assuming a formation factor value of 4.8, or a larger value possibly of 10, will only differ by 1 per cent. This error has been added in Table 1. The $\zeta$-potential decreases from $-16$ to $-27 \pm 4$ mV as the pH increases from 4 to 7. The $\zeta$-potential is constant at $-25 \pm 1$ mV from pH 8 to 10.5. For solution pH higher than 10.5, the $\zeta$-potential is seen to increase in a less regular manner. The $\zeta$-potential of sample 1 (experiment 7 at pH 10.7) is equal to $-11$ mV. While the $\zeta$-potential is negative for pH ranging from 4.4 to 11, a change of sign is observed with sample 2 at pH 11.7 (experiment 12) and the $\zeta$-potential rises to $+8$ mV. At pH higher than 9, there is a decrease in the absolute value of $\zeta$-potential, which even leads to a change of sign. This behaviour is thought to be related to the precipitation of calcite.

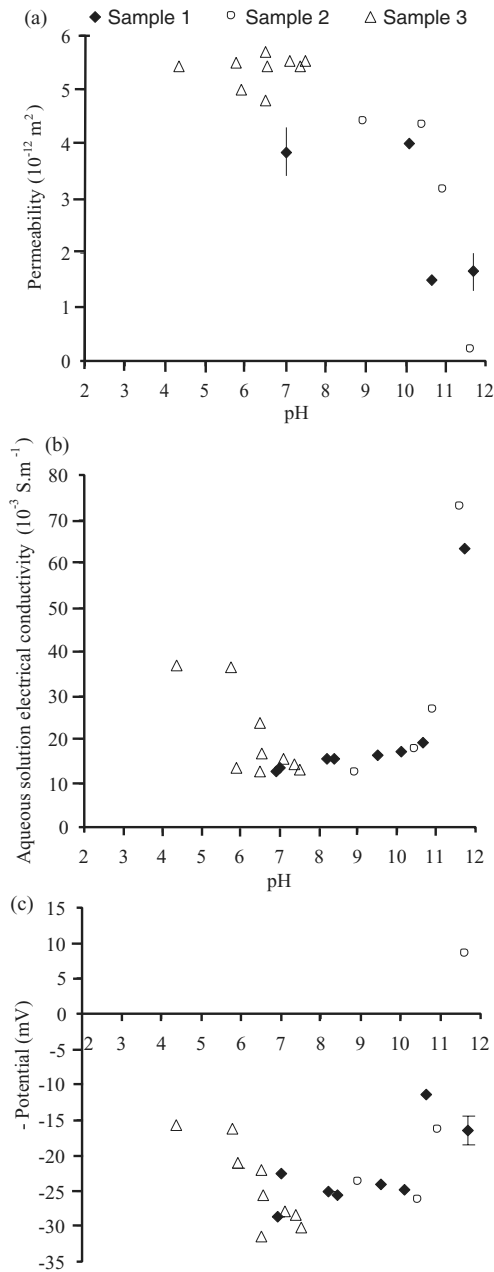

**Figure 5.** (a) Variation of the permeability of the samples as a function of pH (b) Variation of the electrical conductivity of the aqueous solutions, after flowing through the sand, as a function of pH (c) Variation of the $\zeta$-potential derived from eq. (1) using $F = 4.8$, $\sigma_s = 10^{-4}$ S m$^{-1}$ and $m = 1.3$, as a function of pH (d) Variation of the ionic strength of the aqueous solutions as a function of pH calculated using the chemical analyses and WATEQ program (Plummer *et al.* 1976). The ionic strength is not calculated for experiments run at pH higher than 11. Sample 1 was used with pH solutions from 7 to 11.7; sample 2 was used with ph solutions from 9 to 11.7; sample 3 was used with pH solutions from 7 to 4.4.

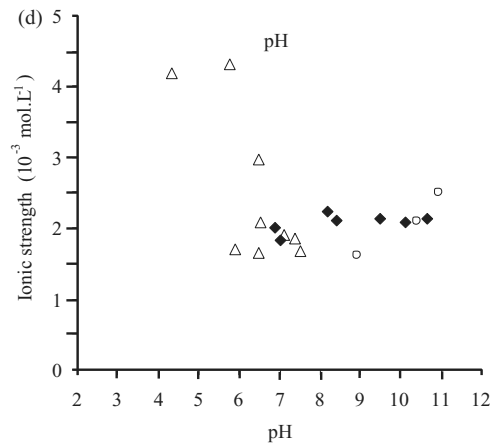

**Figure 5.** (*Continued.*

## 3.2 Chemical analysis

Table 2 groups together the chemical analyses of the aqueous solutions. Since the initial solution is obtained by dissolving CaCl$_2$ in distilled water, the major cation is Ca$^{2+}$ and the major anion is Cl$^-$. The concentration of sodium is two orders of magnitude smaller than the concentration of calcium. We performed speciation calculations with the WATEQ program (Plummer *et al.* 1976), using the revised thermodynamic database of (Ball & Nordstrom 1991). The calculated aqueous speciation for calcium are shown in Fig. 6. For pH lower than 9, calcium does not form any complex in solution. In the pH range 9–12, the WATEQ calculations indicate the appearance of a CaCO$_3^0$ complex, which represents about 10 per cent of the calcium species. These speciation calculations are used to calculate $\zeta$-potentials using the TLM model (Appendix B).

## 4 DISCUSSION

We now discuss the results, in particular the behaviour of zeta potential as a function of pH. We propose to interpret the measured $\zeta$-potentials change of sign for pH larger than 11 by the calcite precipitation, using calculations of the surface electrical potential for both a quartz–water and a calcite–water interface.

### 4.1 Electrokinetic potentials

The ionic strength remains constant in the pH range 5.5–10.7 (Fig. 5d). Fig. 7 shows the evolution of $\zeta$-potential at constant ionic strength ($I = 2 \pm 0.2 \times 10^{-3}$ mol l$^{-1}$) as a function of pH, that is, all experiments except n°s 8, 12, 17, 18 and 19 (Table 1). For comparison measurements from Lorne *et al.* (1999a) and Ishido & Mizutani (1981) are plotted. In the pH range from 4 to 10.5, the shape of the pH vs. $\zeta$-potential dependence of our measurements is in good agreement with the results of Lorne *et al.* (1999a). The absolute values of the measurements reported by Ishido & Mizutani (1981) are larger, and the decrease with pH is much more abrupt. There are several possible reasons for this discrepancy:

(1) Sample preparation: Ishido & Mizutani (1981) carried out measurements with natural quartz that was crushed and cleaned with dilute nitric acid and then washed with distilled water. The samples were stored in distilled water for several months before being used. Lorne *et al.* (1999a) carried out measurements with crushed Fontainebleau sandstone that had not been cleaned with acid.

**Table 2.** Chemical analyses of the aqueous solutions sampled after the experiments. Uncertainties are $0.001 \times 10^{-3}$ mol l$^{-1}$.

| Experiment n° | Ca$^{2+}$ $10^{-3}$ mol l$^{-1}$ | Na$^+$ $10^{-3}$ mol l$^{-1}$ | Cl$^-$ $10^{-3}$ mol l$^{-1}$ | Alkalinity $10^{-3}$ mol l$^{-1}$ | SiO$_2$ $10^{-3}$ mol l$^{-1}$ |
|---|---|---|---|---|---|
| 1 | 0.624 | 0.023 | 1.128 | 0.384 | 0.007 |
| 2 | 0.589 | 0.009 | 1.016 | 0.272 | 0.004 |
| 3 | 0.694 | 0.147 | 1.185 | 0.352 | 0.006 |
| 4 | 0.679 | 0.005 | 0.979 | 0.528 | 0.006 |
| 5 | 0.709 | 0.004 | 1.001 | 0.544 | 0.007 |
| 6 | 0.734 | 0.009 | 0.860 | 0.576 | 0.009 |
| 7 | 0.709 | 0.006 | 1.016 | 0.592 | 0.010 |
| 8 | 0.931 | 0.006 | 0.987 | 0.432 | 0.023 |
| 9 | 0.514 | $ | 0.818 | 0.368 | 0.005 |
| 10 | 0.644 | $ | 0.894 | 0.960 | 0.010 |
| 11 | 0.828 | $ | 0.931 | 0.944 | 0.017 |
| 12 | 1.238 | $ | 0.937 | 1.280 | 0.021 |
| 13 | 0.524 | $ | 0.922 | 0.320 | 0.004 |
| 14 | 0.584 | $ | 1.058 | 0.304 | 0.004 |
| 15 | 0.609 | $ | 1.086 | 0.272 | 0.005 |
| 16 | 0.674 | $ | 1.255 | 0.240 | 0.006 |
| 17 | 0.938 | 0.005 | 1.932 | 0.256 | 0.010 |
| 18 | 1.317 | 0.007 | 2.863 | 0.136 | 0.007 |
| 19 | 1.352 | 0.007 | 3.032 | 0.160 | 0.008 |
| 20 | 0.514 | 0.004 | 1.001 | 0.224 | 0.005 |
| 21 | 0.519 | 0.007 | 1.142 | 0.192 | 0.003 |

$ Below determination threshold.

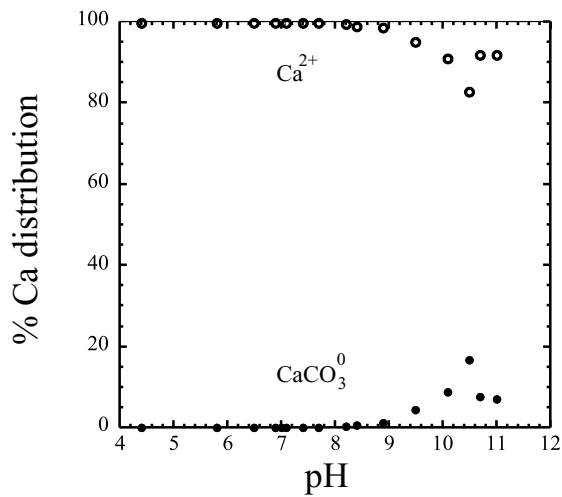

**Figure 6.** Aqueous speciation Ca calculated by WATEQ program (Plummer *et al.* 1976) using chemical analyses from present study (Table 1).

However, Lorne *et al.* (1999a) also ran an experiment with a crushed Fontainebleau sandstone that had been cleaned with hydrochloric acid, showing no significant change in the measured $\zeta$-potential.

(2) Temperature: Lorne *et al.* (1999a) obtained a $\zeta$-potential of about $-25$ mV at pH 6 using NaCl at $10^{-3}$ mol l$^{-1}$, performing their measurements at $22 \pm 1$°C. Ishido & Mizutani (1981) measured $\zeta$-potentials of about $-90$ mV at pH 6 using KNO3 at $10^{-3}$ mol l$^{-1}$, and at a temperature of 45°C. These authors (1981) reported that $\zeta$-potential decreases with increasing temperature, by about $-0.65$ mV/°C at pH 6.1. This would lead to a $\zeta$-potential of about $-75$ mV at 22°C, a value which is still far from the value measured by Lorne *et al.* (1999a).

(3) Permeability: the permeability of our samples is about $5 \times 10^{-12}$ m$^2$ for pH in the range 5.5–10.5. Lorne *et al.* (1999a,b) observed that the $\zeta$-potential remains essentially constant for permeability ranging from $0.01 \times 10^{-12}$ m$^2$ to $10 \times 10^{-12}$ m$^2$, but decreases (in absolute value) at higher permeabilities. The permeability of the samples studied by Ishido & Mizutani (1981) is about $100 \times 10^{-12}$ m$^2$. The discrepancy between our measurements and those of Ishido & Mizutani (1981) may be due to different permeabilities of the samples as noted by Ishido & Mizutani (1981), although the zeta potential is higher (in absolute value).

(4) Equilibrium time: Zeta-potential values deduced by Ishido & Mizutani (1981) were obtained after waiting 5–10 hr, whereas the values deduced by Lorne *et al.* (1999a) were obtained 'when fluid conductivity was stable', the sample could be 'left in contact with the electrolyte for some days', as noted by the authors, without giving any precise equilibrium time. It can not be concluded here that the high values of zeta potential (in absolute value) from Ishido & Mizutani (1981) are due to differences in equilibrium time.

The agreement of our data (using CaCl$_2$ at 0.015 S m$^{-1}$) with those (using NaCl at 0.01 S m$^{-1}$) of Lorne *et al.* (1999a) is quite surprising because $\zeta$-potentials obtained with divalent cations should be twice lower than the values obtained with monovalent cations at similar ionic strength (Lorne *et al.* 1999a). Our measurement performed at pH 5.8 ($-21$ mV) is still higher than the zeta value of $-11$ mV from Lorne *et al.* (1999a) using CaCl$_2$ solution. Moreover, our measurements yield a $\zeta$-potential of $-27 \pm 4$ mV at pH 7, which is about twice lower in absolute terms than the data in the literature for quartz–water interface giving values of about $-70$ mV at an ionic strength of $10^{-3}$ mol l$^{-1}$ (Pride & Morgan 1991), or $-75$ mV (Ishido & Mizutani 1981), which can be considered consistent with the fact that predominant cations are divalent cations in our study.

The measured $\zeta$-potentials fall to $-27$ mV with increasing pH from 4 to 7, but remain constant at $-25$ mV from pH 8 to 10.5. This is surprising because the quartz surface is expected to become more negative with increasing pH, as the number of >SiO$^-$ sites increases. A further reason is that since the concentration of counterions Ca$^{2+}$ is kept constant up to pH 9 (Fig. 6), the $\zeta$-potential should decrease. The aqueous complex CaCO$_3$ appears at about pH 9. Its abundance increases up to pH 10.5 and then stays broadly constant at higher pH. The presence of CaCO$_3$ aqueous complexes decreases the amount of free Ca$^{2+}$ in solution, so the measured $\zeta$-potential should be more negative for pH higher than 8. The constant value of $\zeta$-potential from pH 8 to 10.5 cannot be related to the appearance of aqueous complexes of Ca$^{2+}$.

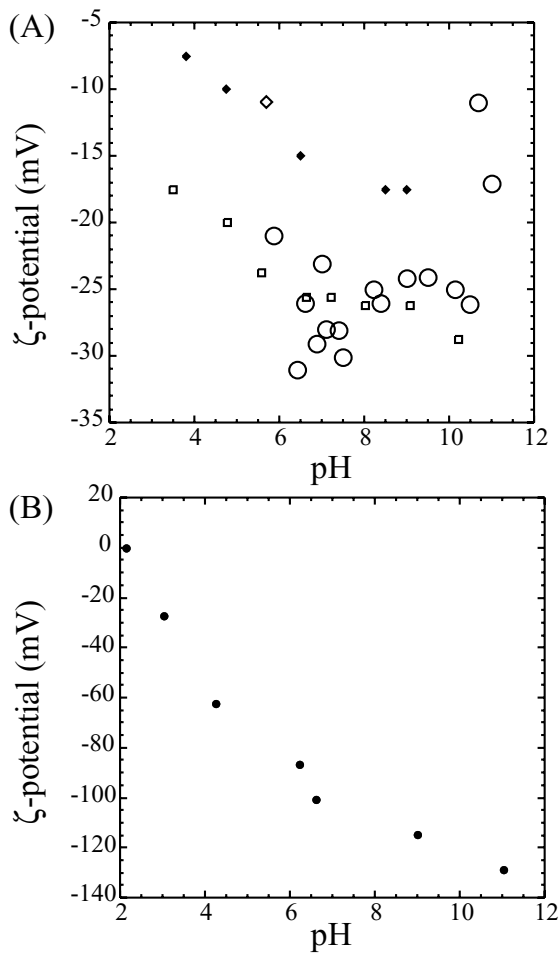

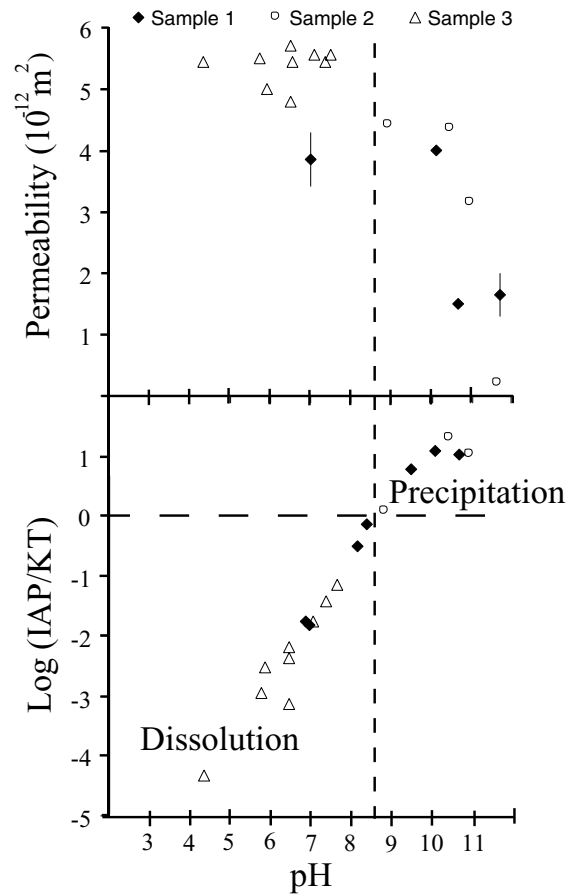

**Figure 7.** (a) Variation of the $\zeta$-potential as a function of pH, at constant ionic strength of the aqueous solution $2 \pm 0.2 \times 10^{-3}$ mol l$^{-1}$ (empty circles). Measurements from Lorne *et al.* (1999a) were performed with a crushed Fontainebleau sandstone, for NaCl aqueous solutions with pH adjusted by adding HCl or NaOH, and for KCl aqueous solutions with pH adjusted by adding HCl or KOH. These measurements were performed with a constant electrolyte resistivity of 100 $\Omega$.m in the case of NaCl (empty squares) solutions (about $10^{-3}$ mol l$^{-1}$) and 25.5 $\Omega$ m (about $2 \times 10^{-2}$ mol l$^{-1}$) in the case of KCl solutions (diamonds). Measurement using CaCl$_2$ solution is also reported (empty diamond): the inferred $\zeta$-potential of $-11$ mV at $pH = 5.7$ for a fluid conductivity 13.5 mS m$^{-1}$ (or 74 $\Omega$ m). (b) Measurements from Ishido & Mizutani (1981) were performed on Ishikawa quartz using aqueous solutions of KNO$_3$ ($10^{-3}$ mol L$^{-1}$), with the temperature set at 45°C.

### 4.2 Calcite precipitation

The measured $\zeta$-potentials are scattered for pH higher than 10.5 (Fig. 7). We assume that this scattering is related to calcite precipitation, as inferred from saturation index calculations and from the decrease in permeability. The values of permeability of sample

**Figure 8.** Saturation indices of calcite as a function of pH, associated with the variation in sample permeability. When the solution is oversaturated with respect to calcite, the permeability of the sample decreases.

1 (experiment 8) and sample 2 (experiment 12), at pH 11.7, are $1.7 \times 10^{-12}$ m$^2$ and $0.2 \times 10^{-12}$ m$^2$, respectively. These two values are smaller than the initial values obtained on sample 1 (sample 2) $4 \times 10^{-12}$ m$^2$ (4.4 $\times 10^{-12}$ m$^2$) at neutral pH. This decrease in permeability is thought to be due to the precipitation of calcite. Using the WATEQ program (Plummer *et al.* 1976) and the chemical analyses of water samples in Table 2, we calculate the saturation indices of the calcite. This index is defined as the logarithm of the ionic activities product (IAP), that is, as the product of the Ca$^{2+}$ activity and the CO$_3^{2-}$ activity, divided by the calcite solubility product. Fig. 8 shows the variation of the calcite saturation index linked to permeability as a function of pH. Since the Ca$^{2+}$ activity remains approximately constant, the increase of the saturation index with increasing pH is related to the increase of CO$_3^{2-}$ activity with increasing pH (Fig. 6). The saturation index is less than 0 when pH ranges from 4.4 to 8.4, that is, the calcite of the sample undergoes dissolution. When pH is higher than 8.7, the saturation index is greater than zero, calcite precipitates from the solution and the permeability of the sample decreases. The calcite precipitate is expected to form a *coating* on the quartz grains. The higher the solution pH, the greater the amount of precipitated calcite, because (i) the saturation index increases, (ii) the duration of the experiment increases since the same sample remains in contact with a basic pH solution. The calcite *coating* gradually fills up the pore space and hydraulic pathways.

### 4.3 Surface complexation model

We performed calculations of the surface electrical properties in order to achieve a semi-quantitative description of our measurements, and to investigate if calcite precipitation can lead to a reversal of sign of the $\zeta$-potential. The zeta potential is first calculated for a quartz–water interface, and then for a calcite–water interface. The surface electrical properties of minerals are described by surface complexation models, which are based on a description of the formation of complexes on the surface of minerals, that is, between surface functional groups and dissolved species in the electrolyte. We choose TLM because consistent parameters of the model (i.e. capacitive constants, complexation constants) are available for numerous minerals. A mathematical description of TLM and a review of the quartz and calcite surfaces properties are given in Appendix B. The concentrations used for run calculations are the output species concentrations obtained from the WATEQ program (Plummer *et al.* 1976; Ball & Nordstrom 1991), using the results from the chemical analyses of water samples (Table 2). Equilibrium constants and capacitance constants are not adjustable parameters, but are values taken from the literature, and the sensitivity of the different possible values of these parameters on zeta-potential calculations has been achieved (see details in Appendix B and Fig. 9).

**Quartz–Water Interface:** Calculations were performed using the surface complexation reactions (Table 3), $C_1 = 1.4$ F m$^{-2}$ or 0.81 F m$^{-2}$, $C_2 = 0.2$ F m$^{-2}$ (see Appendix B for details), and total surface functional group densities of 10 and 25 sites nm$^{-2}$ for pH in the range 4–8 and for pH higher than 8, respectively. Results of TLM calculations for a quartz–water interface are compared to our measurements in Fig. 9A. The calculated $\zeta$-potentials decrease with increasing pH from 4.4 to 7.7, and remain constant in the pH range 8–10.5. The calculated $\zeta$-potentials are in a good agreement with the measured $\zeta$-potentials for pH ranging from 5.5 to 10. At pH 4.4, the calculated $\zeta$-potential (around $-6$ mV) is higher than the measured $\zeta$-potential ($-16$ mV). At pH higher than 10.5, the calculated $\zeta$-potentials at a quartz–solution interface do not agree with the measured values.

**Calcite–Water Interface:** Calculations were performed using the surface complexation reactions (Table 4), $C_1 = 1.4$ F m$^{-2}$, $C_2 = 0.2$ F m$^{-2}$, and total surface functional group densities of 5 sites.nm$^{-2}$ (see Appendix B for details). Results of TLM calculations for a calcite–water interface are compared to our measurements in Fig. 9B. Let us notice that the calculated $\zeta$-potentials show a broad range of values depending on equilibrium constants published in recent geochemical literature (Van Cappellen *et al.* 1993; Pokrovsky *et al.* 1999; Fenter *et al.* 2000). Therefore, we are only able to provide a semi-quantitative interpretation of our measurements according to TLM results. TLM calculations do not agree with our measurements while pH increases from 4.4 to 10.5. The measured $\zeta$-potentials show large variations and change of sign for pH higher than 10.5 (Fig. 7a). However, the magnitude of these variations can be compared with the magnitude of the range of TLM calculations values (Fig. 9B). TLM calculations also present a change of sign; electrical surface potentials calculated according to Van Cappellen *et al.* (1993)'s (or Pokrovsky *et al.* 1999's) equilibrium constants are always positive, whereas electrical surface potentials calculated according to Fenter *et al.* (2000)'s equilibrium constants became negative for pH larger than 8.

In summary, our $\zeta$-potential measurements can be explained by a quartz–water interface when pH ranges from 5.5 to 8.5, and by a calcite–water interface when pH is greater than 10.5. When pH ranges from 8.5 to 10.5, the measured $\zeta$-potentials are higher than the $\zeta$-potentials calculated for a quartz–water interface and lower than the $\zeta$-potentials calculated for a calcite–water interface. These measured $\zeta$-potential values are probably representative of a surface composed by a mixture of quartz and calcite. Note that constant zeta potentials in the pH range 8.5 to 10.5 do not reflect the quartz–water interface usually described in the literature [Pride and Morgan, 1991], which shows a large decrease in zeta potential with increasing pH (see Fig. 7b). Our predicted values are flat in this pH range since our calculations are made using the [$Ca^{2+}$] measured in our experiments. In this pH range, the predicted values for quartz–water interface are closer to the measurements than the predicted values for calcite–water interface, probably because of the unfinished coating of calcite on quartz grains since it is only the beginning of calcite precipitation (although it was not possible to quantify the calcite precipitation). The measured $\zeta$-potentials change of sign for pH greater than 11, can be explained by the calcite precipitation inferred from the measured decrease of permeability and from chemical speciation calculations.

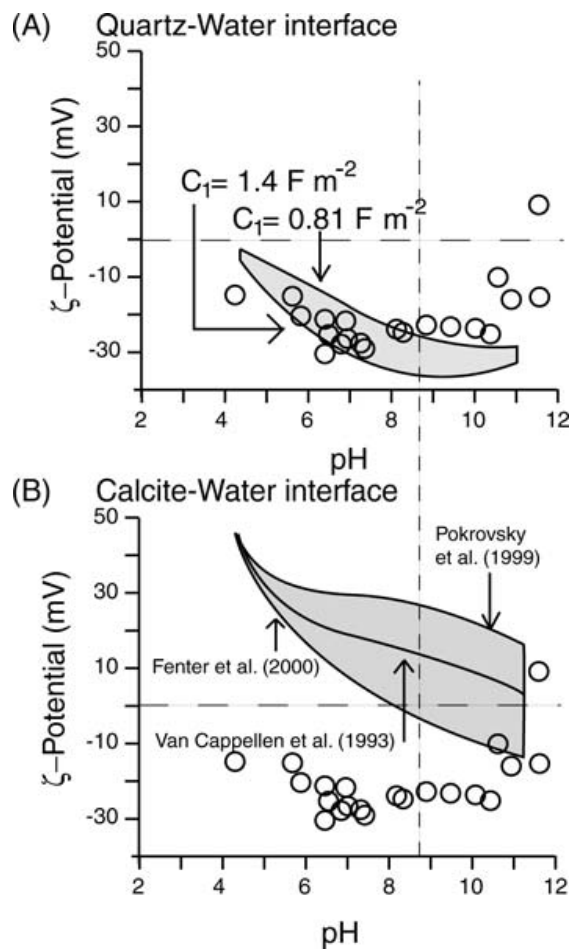

**Figure 9.** Results of the TLM calculation for (a) quartz–water interface, (b) calcite–water interface. The shadow parts show the sensitivity of the possible values of the electrical surface potentials as a fonction of input parameters (see Appendix B for details). The vertical dashed line is deduced from Fig. 8, and indicates the limit of oversaturation with respect to calcite (for pH greater than 8.7, the solutions are oversaturated with respect to calcite). The chemical surface reactions considered are grouped in Tables 3 and 4.

**Table 3.** Surface complexation reaction and equilibrium constants for the quartz–water interface used in the triple-layer model.

| Surface reaction | Log $K$ | Adsorption plane |
|---|---|---|
| $> \text{SiOH H}^+ \rightleftarrows > \text{SiOH}_2^+$ | $-1.3^{(a)}$ | IHP |
| $> \text{SiO}^- + \text{H}^+ \rightleftarrows > \text{SiOH}$ | $7.2^{(a)}$ | IHP |
| $> \text{SiO}^- \text{Ca}^{2+} \rightleftarrows > \text{SiO}^- \text{Ca}^{2+}$ | $-1.19^{(b)}$ | $\beta$-plane |
| $> \text{SiOH}_2^+ \text{Cl}^- \rightleftarrows > \text{SiOH}_2^+ - \text{Cl}^-$ | $0.55^{(a)}$ | $\beta$-plane |

$^{(a)}$From Sverjensky & Sahai (1996) and Sahai & Sverjensky (1997b).
$^{(b)}$Calculated using equations (24), (35) and (36) from Sahai & Sverjensky (1997b).

**Table 4.** Surface complexation reaction and equilibrium constants for the calcite–water interface used in the triple-layer model.

| Surface reaction | Log K | Reference |
|---|---|---|
| $> CO_3H^0 \rightleftarrows > CO_3^- + H^+$ | −4.9 | Van Cappellen *et al.* (1993) |
|  | −2.8 | Van Cappellen *et al.* (1993) |
| $> CO_3H^0 + Ca^{2+} \rightleftarrows > CO_3Ca^+ + H^+$ | −4.4 | Fenter *et al.* (2000) |
|  | −1.7 | Pokrovsky & Schott (1999) |
| $> CaOH_2^+ \rightleftarrows > CaOH^0 + H^+$ | −12.2 | Van Cappellen *et al.* (1993) |
| $> CaOH^0 \rightleftarrows > CaO^- + H^+$ | −17 | Van Cappellen *et al.* (1993) |
| $> CaOH^0 + CO_2 \rightleftarrows > CaHCO_3^0$ | 6 | Van Cappellen *et al.* (1993) |
| $> CaOH^0 + CO_2 \rightleftarrows > CaCO_3^- + H^+$ | −2.6 | Van Cappellen *et al.* (1993) |

### 4.4 Electrical surface conductivity

We now discuss the sensibility of the inferred $\zeta$-value to the surface conductivity value, as described by eq. (1). When calcite precipitates, it coats the quartz surface. Therefore, the electrical surface conductivity of the sand with a calcite coating is expected to be different than the electrical surface conductivity of the initial sand. To estimate the new electrical surface conductivity of the sand with calcite coating, we should measure the electrical conductivity of the sample versus the electrical conductivity of the fluid saturating the sample, for a set of fluid electrical conductivities ranging from demineralized water up to saline water (about $1 \, S \, m^{-1}$). However, if the sample is saturated with demineralized water, the coating is expected to be dissolved. Therefore, to check the sensitivity of the $\zeta$-potentials values deduced from eq. (1) with an electrical surface conductivity representative of carbonated rocks, we calculate $\zeta$-potentials, for pH greater than 10, according to eq. (1) with a surface electrical surface conductivity equal to $1.4 \pm 0.5 \times 10^{-3} \, S \, m^{-1}$ (instead of $1.06 \times 10^{-4} \, S \, m^{-1}$), which is the mean value of 14 electrical surface conductivities of carbonated rocks (M. Zamora, personal communication) (Fig. 10). The zeta potential is enhanced (in absolute value) by $\sim$10 to 40 per cent, showing that it is important to know precisely this parameter. However, the general trend and obviously the change of sign observed in electrokinetic measurements (Fig. 10) are the same within our experimental errors.

### 4.5 Consequences for field interpretations

It has been shown that large hydrothermal system without prominent surface area, and probably related to deeper intra-edifice system, can be recognized only through geophysical and geochemical measurements (Finizola *et al.* 2003). In order to interpret the SP surveys the streaming potential is always assumed positive in the flow direction. We would like to emphasize that this assumption is not correct when the zeta potential is positive instead of usually negative and we explore the field conditions needed to obtain a negative charge separation along the fluid flow.

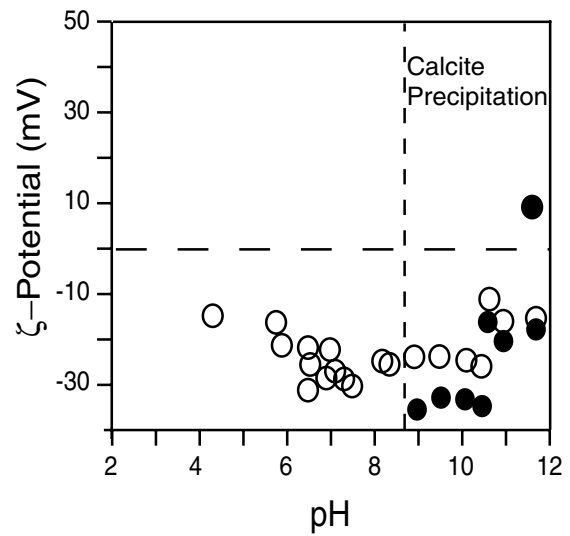

**Figure 10.** Check of the sensitiveness of $\zeta$-potential values for various electrical surface conductivities when pH is greater than 8.7. $\zeta$-potential deduced from eq. (1) using a surface electrical conductivity of $1.06 \times 10^{-4} \, S \, m^{-1}$ (empty circles), and deduced from eq.(1) using a surface conductivity of $1.4 \times 10^{-3} \, S \, m^{-1}$. The latter surface electrical conductivity is representative of carbonated rocks.

The distribution of SP and $CO_2$ anomalies observed on volcanoes is greatly influenced by the presence of structural limits such as crater, caldera or landslide faults. These discontinuities clearly constitute permeable drains on Stromboli (Finizola *et al.* 2003) allowing the escape of upward hot fluids and gas from the hydrothermal system: the electrical charge separation being usually positive in the flow direction, a positive SP anomaly is usually associated to a structural limit. In the summit area of Stromboli particularly a very dense survey was achieved for the first time in a similar context (Finizola *et al.* 2003), and the co-occurrence of SP, temperature and $CO_2$ positive anomalies above the main faults lead to the conclusion that the transfer of hot hydrothermal fluids, heat and gases around the active area was controlled by the presence of more permeable paths along the fault. It was also observed a 'cold zone' interpreted as the filling of the craters by highly impermeable products, fluids, and heat only escaping along the border faults of the craters. In contrast, further from the active area, a previous study (Finizola *et al.* 2002) showed an inverse relationship between SP and $CO_2$ anomalies at the scale of the island of Stromboli. High $CO_2$ emanations were correlated with significant negative SP anomalies. This correlation was interpreted considering the permeable faults away from the active zone still allowing the upward migration of deep $CO_2$ degassing, but promoting the down-ward migration of ground water. This interpretation is in agreement with the assumption that charge separation is positive in the flow direction, but would be exactly the reverse assuming a reverse streaming potential. Moreover the occurrence of high-frequency SP signal closely associated with temperature anomalies, therefore, associated with high hydrothermal flux condensing close to the surface (Finizola *et al.* 2003) may have been interpreted considering the possibility of alteration and apparition of secondary minerals displaying a reverse zeta potential.

These results show the limits of the usual qualitative interpretation of the small-scale pattern of SP anomalies in order to determine local fluid circulations.

We now explore the field conditions needed to obtain negative charge separation along a fluid flow, meaning a positive zeta potential. We showed that this reverse sign in the streaming potential can be observed when secondary minerals such as calcite are present, for high pH values. These characteristics may not be evident to be observed on the hydrothermal systems of active volcanoes showing mixing between acid gas species such as $H_2S$, $SO_2$, HCl, HBr, HF, which lead to acid pH. However, calcite is often present in hydrothermal systems as an alteration product of volcanic rocks, precipitating on the walls of the rock cavities (Keller *et al.* 1979; Henley & Ellis 1983; Robert 2001). Evidences of high hydrothermal flux condensing close to the surface can be inferred by soil temperature elevation performed in the upper part of active volcanoes. These thermal anomalies are thought to coincide with permeable zones where uprising hot fluids escape (Finizola *et al.* 2003), leading to alteration and possibly to secondary minerals. Since electrical surface potential of calcite can range from positive to negative values depending on $CO_2$ partial pressure, pH, and $[Ca^{2+}]$ (Pokrovsky *et al.* 1999), either SP maxima or SP minima can be correlated to $CO_2$ maxima with the same fluid circulation direction even at pH lower than 9. Moreover calcite precipitation can also be effective on active volcano in acid environment: actually, in complex volcanic system, with a large output of magmatic $CO_2$ and low contribution of HCl, calcite can precipitate at low temperature such as $100°C–130°C$ or at higher temperature such as $>300°C$ in a range of pH of 4.5–5 (Di Liberto *et al.* 2002).

## 5 SUMMARY AND CONCLUSION

The electrokinetic properties of the studied quartz–calcite sand are significantly affected when precipitation of calcite as a secondary mineral is observed. The $\zeta$-potential is greatly reduced (in absolute value) and can even change sign. The electrical surface potentials calculated using TLM as a function of the pH, indicate that the electrical surface potential values are controlled by a quartz–water interface when pH ranges from 5.5 to 8.5, and by a calcite–water interface when pH is greater than 10.5.

SP method is frequently used on active volcanoes to evidence the hydrothermal systems and to outline their extension (Michel & Zlotnicki 1998; Lénat *et al.* 2000; Finizola *et al.* 2002, 2003, 2004). The small-scale patterns of SP and $CO_2$ anomalies are used to study tectonic fracture and structural limits. The electrokinetic coupling is always assumed to be positive in the flow direction—therefore, the negative anomalies are interpreted in terms of downward percolating rainfall, and the positive anomalies originate from upward convective flows (Zablocki 1976, 1978). This assumption is accurate if the surface electrical potential of the rocks is negative (see eq. 1). Surprisingly the SP studies do not check the nature of the rocks, whereas the surface electrical potential of the rock can be positive sometimes. In the field, only one experiment has been performed controlling $CO_2$ injection while measuring SP and showed an inversion in SP signals compared to the injection of neutral gas (Martinelli 2000). Calcite is often present in hydrothermal systems as an alteration product of volcanic rocks, precipitating on the walls of the rock cavities (Keller *et al.* 1979; Henley & Ellis 1983; Robert 2001). Since electrical surface potential of calcite can range from positive to negative values depending on $CO_2$ partial pressure, pH, and $[Ca^{2+}]$ (Pokrovsky *et al.* 1999), either SP maxima or SP minima can be correlated to $CO_2$ maxima with the same fluid circulation direction even at pH lower than 9. Moreover calcite precipitation can also occur in acid environment such as pH of 4.5–5 on active volcano with a large output of magmatic $CO_2$ and low contribution of HCl (Di Liberto *et al.* 2002).

Therefore, SP anomalies as measured in hydrothermal field, without mineralogical analyses of hydrothermal deposits, and without geochemical fluid survey, should be interpreted with caution, to infer fluid circulations.

## ACKNOWLEDGMENTS

We thank G. Marolleau for the construction of the apparatus. We thank O. Pokrovsky, J. Schott, C. Robert for constructive remarks. We thank A. Finizola and T. Ishido for their constructive reviews. This research was supported by CNRS and by ACI- Prévention des Catastrophes Naturelles of the French Ministry of Research. Dr M. S. N. Carpenter carried out post-editing of a previous version of this final manuscript.

## APPENDIX A: EQUILIBRIUM TIME

We detail here how we checked that the equilibrium between the rock and the electrolyte was attained for sample 1. Measurements of electrokinetic coupling, pH, and fluid conductivity as a function of time are shown in Fig. A1. The initial water ($CaCl_2$ solution) has a conductivity 0.0127 S m$^{-1}$ and pH 5.5. After 750 mL flowed through the sample, the conductivity and the pH of the output water are $\sigma_f = 0,0116$ S m$^{-1}$ and $pH = 6.8$. Then we wait 21 hr before the output water is made to flow through the sample. The water is circulated twice 750 mL to check if the conductivity and pH are constant: at this time, if the conductivity and pH of the output water are constant, equilibrium is thought to be attained and streaming potential as a function of various applied pressures is measured. The linearity between the streaming potential and the applied pressure is checked and the electrokinetic coupling coefficient is deduced from 2 to 3 sets of measurements (see for examples Fig. 4). The measurements performed at time 22 hr are made with the initial solution (exp. 1), whereas measurements performed at time 44 hr (exp. 2) are performed using the initial solution added with $CaOH_2$, although the conductivity and pH are almost the same. These measurements lead to a $\zeta$-potential of −29 and −23 mV, respectively. At time 70 hr the pH of the solution is increased up to 8.9 by adding $CaOH_2$. Electrokinetic measurements are performed after waiting 1 hr and the results show a large dispersion, probably because the equilibrium is not attained. These measurements were not taken into account. After waiting 20 hr (time 90 hr) the output water is made to flow trough the sample and the conductivity and pH of the ouput water are $\sigma_f = 0,0155$ S m$^{-1}$ and $pH = 8.3$. This water is recirculated twice in the sample and the conductivity and pH are constant: the electrokinetic coupling coefficient can be, therefore, quantified (exp. 3 and 4). Then the pH of water is increased up to 9.5 and after waiting 27 hr the output water is recirculated twice and the conductivity and pH are checked to be constant and the electrokinetic coupling coefficient is quantified (exp. 5). All the measurements have been performed using this experimental protocol.

The long run times needed to reach chemical equilibrium have already been emphasized by Davis & Kent (1990), who questioned the applicability of surface complexation models to carbonates. Furthermore, Zachara *et al.* (1993) observed that surface adsorption processes for divalent cations onto calcite and carbonates occur over two characteristic durations: one of a few hours, in relation to reversible reactions, and another of several days associated with irreversible reactions. The latter phenomena involve changes in the chemical bonding of the adsorbed divalent cations, inducing the formation of a recrystallized surface phase or a solid solution. The latter phenomena are not investigated in the present paper.

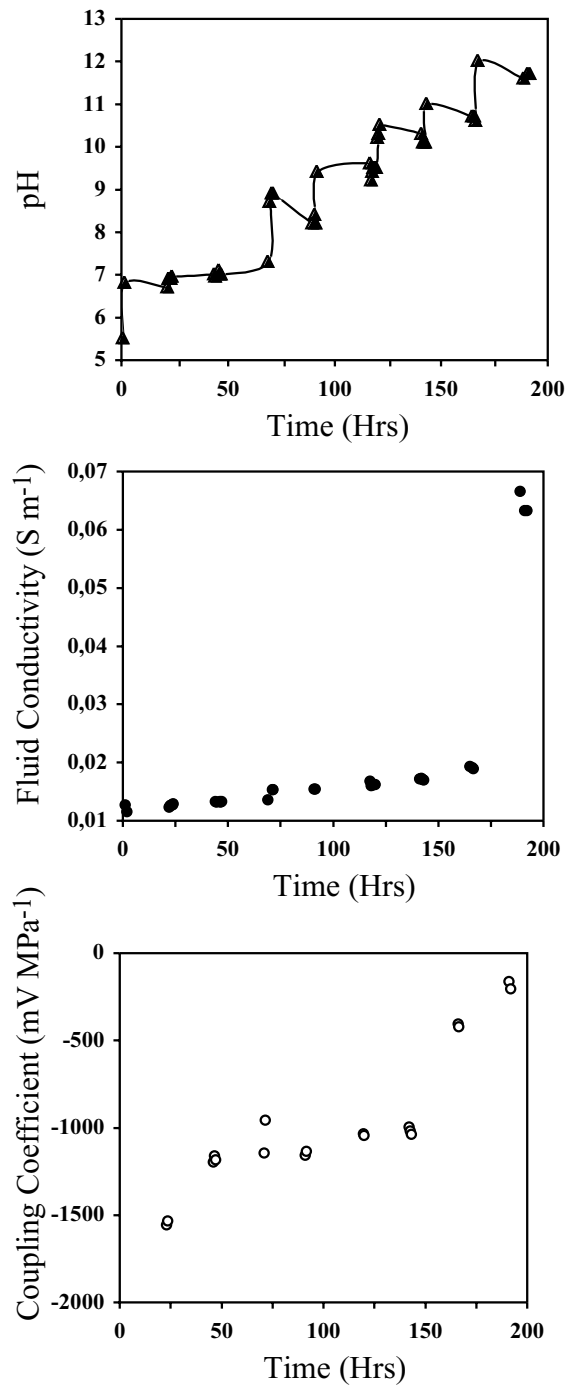

**Figure A1.** Measurements of the electrokinetic coupling, pH, and fluid conductivity, as a function of time.

**Error sources**

The error on each zeta potential deduced from eq. (1) was calculated by the sum of the errors on: the dielectric constant, the water conductivity, the viscosity, and the measured coupling coefficient. Temperature variation is taken into account for the dielectric constant. The error on fluid conductivity is ±0.4 per cent; the error on viscosity due to temperature variations is for most of the experiments between 1 and 5 per cent, but is between 9 and 13 per cent for the experiments performed a temperature below 16.5°; the error on the electrokinetic coupling coefficients is ±1 to ±5 per cent for most of the measurements, and ±12 per cent for exp. 8. Each error has been quantified on zeta potential (Table 1) and is included in the size of drawing in Figs 5, 7, 9 and 10. The surface conductivity is taken into account, however, since the exact value is not known precisely for pH higher than 9, this point has been developed in Section 4.4.

The measurements performed at time 22 hr are made with the initial solution (exp. 1) and lead to a zeta potential of −29 mV. The measurements performed with the initial solution on two other samples (exp. 13 and 21) lead to a zeta potential of −30 and −31 mV, respectively. Therefore, the variation from sample to sample leads to an error of ±3.5 per cent, which is, for most of the measurements, the same order of magnitude of the error due to the eq. (1).

## APPENDIX B: $\zeta$-POTENTIAL MODELLING ACCORDING TO TLM

We describe here the TLM, which is used to calculate zeta potentials of our sand/calcite–water interface to interpret the electrokinetic measurements as a function of pH (see Section 4).

The TLM distinguishes three planes to describe the electrical double layer. One plane, called the Inner Helmholtz Plane (IHP), for counter ions that are directly bound to the mineral structure and another plane, called the Outer Helmholtz Plane (OHP), for weakly bound counter ions (see Fig. B1) (Grahame 1947). The counter ions of the IHP are assumed to be chemically adsorbed, forming a chemical bond with the surface functional group. On the other hand, the counter ions of the OHP are assumed to be physically adsorbed and more labile. The TLM also takes account of a third plane, called the *d*-plane, associated with the smallest distance between the mineral surface and the counter ions in the diffuse layer. At every plane is related an electrical surface charge density $\sigma$ and an electrical potential $\phi$. The electrical surface charge densities and potentials are supposed to follow the Stern–Grahame relationships (Grahame 1947):

$$\phi_0 - \phi_\beta = \sigma_0/C_1$$
$$\phi_\beta - \phi_d = (\sigma_0 + \sigma_\beta)/C_2$$
$$\sigma_0 + \sigma_\beta + \sigma_d = 0 \tag{B1}$$

where the subscript 0, $\beta$ and $d$ refer to IHP, OHP, and *d*-plane, $C_1$ and $C_2$ the integral capacitances of the interfacial layers. The electrical potential on the *d*-plane is usually estimated by the Gouy–Chapman charge–potential relationship based on the bulk solution chemistry:

$$\sigma_d = 2\chi_d \sum_i eZ_i[i] \exp\left(\frac{-q_i\phi_d}{2kT}\right), \tag{B2}$$

where $Z_i$ is the electrical charge of the *i*th ion in solution, $e$ is the elementary electron charge, $k$ is the Boltzmann's constant, $T$ the absolute temperature, and $\chi_d$ is the Debye length equal to:

$$\chi_d = \sqrt{\varepsilon kT/e^2 \sum_i Z_i^2[i]}. \tag{B3}$$

The reader can find a comprehensible report of uses and results of the TLM in Davis & Kent (1990).

The Stern–Grahame model assumes that the surface of the mineral, the $\beta$-plane and the *d*-plane can be considered as plates of a planar condenser. The model of a planar condenser is appropriate if the thickness of the diffuse layer, around two Debye lengths, is small compared with the grain radius. In our study, the mean grain diameter is $3 \times 10^{-4}$ m, and the Debye length is smaller than $10^{-9}$ m.

It has been proposed that the approximation be made that the slipping plane lies near the distance of closest approach of dissociated ions, that is, $\phi_d = \zeta$ (Davis & Kent 1990). The surface electrical potential of the *d*-plane can be determined by solving eqs (B2) and (B3). In order to solve these equations, the electrical surface charge densities $\sigma_0$ and $\sigma_\beta$ must be determined from the surface complexation reactions, which themselves depend on the surface functional groups and on the ionic species in the aqueous solution.

**Quartz surface**

We describe here the surface functional group used to model the quartz surface, and we detail the appropriate values of the total surface functional group density: for pH range 4–8 and for pH greater than 8 we use a total surface functional group density of 10 sites nm$^{-2}$, and 25 sites nm$^{-2}$, respectively.

The quartz surface has been extensively studied, and its properties can be modelled with silanol >SiOH group (Davis *et al.* 1978; Sahai & Sverjensky 1997a,b). Many experiments have been performed with silica gels (Ahrland *et al.* 1960; Dugger *et al.* 1964) and with natural silica (Somasundaran & Kulkarni 1973; Schindler *et al.* 1976; Ishido & Mizutani 1981; Lorne *et al.* 1999a). In summary, the OH$^-$ and H$^+$ ions, termed potential-determining ions, are adsorbed onto the mineral surface and determine the charge density $\sigma_0$. Thus, the surface charge of quartz is a function of pH; the charge is positive for pH < pzc and negative for pH > pzc. The pzc of quartz is in the range 2 < pH < 4 (Parks 1965; Tadros & Lyklema 1969; Parks 1984; Glover *et al.* 1994; Lorne *et al.* 1999a). Since the pH of natural waters ranges from 4 to 10 (Sigg *et al.* 2000), the quartz surface is generally negatively charged. The major cations of natural waters ($Ca^{2+}$, $Mg^{2+}$, $Na^+$ and $K^+$) are found to be adsorbed onto the $\beta$-plane. Two reactions are possible with divalent cations (Ahrland *et al.* 1960; Dugger *et al.* 1964; Tadros & Lyklema 1969; Schindler *et al.* 1976; Stumm *et al.* 1976); the cation can form either a monodentate or a bidentate surface complex.

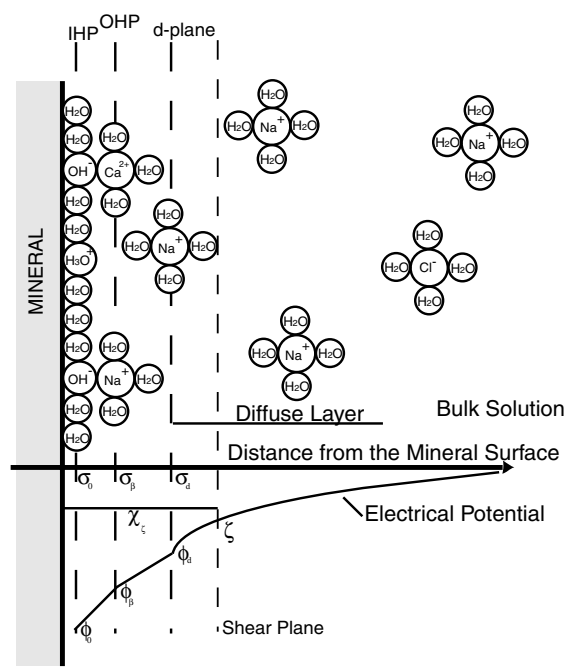

**Figure B1.** Schematic view of the mineral–water interface for oxide minerals, showing location of IHP and OHP, and the potential decay away from the surface (after Davis *et al.* 1978).

The surface parameters used for quartz in the present model are obtained from (Sahai & Sverjensky 1997a,b; Sverjensky & Sahai 1996) and are listed in Table 1. Since $Ca^{2+}$ is the major ion during the experimental runs (Table 1), we ignore the other species such as $Mg^{2+}$, $Na^+$ and $K^+$. To simplify the calculation, we do not consider the formation of a bidentate surface complex of calcium. For pH in the range 4–8 and for pH higher than 8, we use total surface functional group densities of 10 and 25 sites.nm$^{-2}$, respectively.

### Calcite surface

We describe here the surface functional group used to model the calcite surface, and we show the different possible values of the calcification reaction constants.

The surface electrical potential of calcite has been investigated by many geochemists over the last 50 yr (Douglas & Walker 1950; Somasundaran & Agar 1967; Smallwood 1977; Foxall *et al.* 1979; Siffert & Fimbel 1984; Thompson & Pownall 1989; Cicerone *et al.* 1992; Vdovic 2001) and measurements are sometimes contradictory (Fig. 2). For instance, the zero point of charge (pH$_{pzc}$) varies according to authors from 7 to 10.8 (Van Cappellen *et al.* 1993). Curves of $\zeta$-potential versus pH show hysteresis loops (Thompson & Pownall 1989), while different electrokinetic behaviours are observed on natural or synthetic calcite, the $\zeta$-potentials of natural calcite being systematically lower than the values obtained for synthetic calcite (Vdovic 2001).

In summary, $Ca^{2+}$ and carbonate ions are the potential-determining ions (Somasundaran & Agar 1967; Foxall *et al.* 1979; Cicerone *et al.* 1992). $H^+$ and $OH^-$ modify the surface electrical potential because they regulate the concentrations of $Ca^{2+}$ and carbonate ions. Contradictory behaviours are observed at a given pH, because (i) the concentrations of the potential-determining ions are variable (Thompson & Pownall 1989; Cicerone *et al.* 1992); (ii) important dissolution and precipitation reactions imply that chemical equilibrium cannot be reached easily (Somasundaran & Agar 1967), and that—in the presence of other ions—the rapid exchange rates can transform the nature of the solid surface layer (Fuller & Davis 1987; Cicerone *et al.* 1992; Zachara *et al.* 1993). The most common approach to evaluate the validity of surface complexation theory is through measurements of surface charge (Fenter *et al.* 2000). However, calcite surface charge variation versus pH is difficult to measure by potentiometric titrations because of calcite's rapid dissolution (Charlet *et al.* 1990).

Therefore, different values of the calcification reaction constants are published in the literature. Van Cappellen *et al.* (1993) proposed a model with two surface functional groups, $>CaOH^0$ and $>CO_3H^0$, which can react with $Ca^{2+}$, carbonate ions, $H^+$ and $OH^-$ (Table 4). The surface functional groups are based on X-ray photoelectron spectroscopic observations measured under ultra-high vacuum conditions (Stipp & Hochella 1991). The equilibrium constants are calculated to obtain a pH$_{ZPC}$ near 8.2 with atmospheric $CO_2$ partial pressure. This model predicts that the calcite surface has distinct termination depending on solution composition (pH, pCa, and $P_{CO2}$) and shows a reaction constant $\log K = -2.8$ (Table 4). Fenter *et al.* (2000) present *in situ* X-ray reflectivity measurements of the calcite–water interface at pH ranging from 6.8 to 12.1 and low $P_{CO2}$ which show that the calcite surface does not vary significantly over this range of experimental conditions, and can be explained without invoking changes other than protonation reactions in the surface speciation. Particularly Fenter *et al.*'s measurements do not require a calcification reaction as proposed by Van Cappellen *et al.* (1993): $>CO_3H^0 + Ca^{2+} \rightleftharpoons >CO_3Ca^+ + H^+$, and a value of $\log K < -4.4$ satisfies X-ray reflectivity measurements. On the other hand, Pokrovsky & Schott (1999) provide an higher estimate for the $\log K$ ($-1.7$) of the calcification reaction.

Finally, we use Van Cappellen's model, and the variability of calcification reaction constants on the calculated zeta potentials is taken into account (Fig. 9), using a total surface site density value of 5 sites per nm$^2$ as reported by Davis & Kent (1990).

### Integral Capacitances of the interfacial layer

The values of the integral capacitances of the interfacial layers, $C_1$ and $C_2$, are related to the dielectric properties of the interface, particularly the water permittivity value. Despite the widespread application of such surface complexation models, the choice of the integral capacitance values is not straightforward because these values depend on the specific oxide and on the type of electrolyte, and cannot be directly measured (Sverjensky 2001). Usually $C_1$ and $C_2$ values are chosen as adjustable parameters, consistent with interfacial dielectric constant 80, and equal to 1–1.4 and 0.2 F m$^{-2}$, respectively (Davis & Kent 1990). Although measurements of the water permittivity at the interface are scarce and difficult, the water permittivity is supposed to be smaller near the mineral surface than within the bulk solution, <10 (e.g. Kurosaki 1954). Recently Sverjensky (2001) proposed a model consistent with interfacial dielectric constants ranging from 20 to 62, and obtained $C_1 = 0.81$ and 1.00 F m$^{-2}$ for quartz with $CaCl_2$ electrolyte and NaCl electrolyte, respectively.

Calcite, and more generally carbonate minerals, develop relatively high surface charge, about 100 times higher than oxides, and show $\zeta$-potential values of the same order of magnitude (Pokrovsky *et al.* 1999). Van Cappellen *et al.* (1993) propound that carbonate-water interface is thin and has highly structured (i.e. non-diffuse) electric double layer with high capacitance. It is likely, in the case of calcite, that $Ca^{2+}$ and $HCO_3^-$ ($CO_3^{2-}$) form inner-sphere specific interactions with the charged surface, leading to the formation of a very thin double layer (Pokrovsky *et al.* 1999). This could account for great difference in composition and structure of the EDL formed by oxides or by carbonates (Pokrovsky *et al.* 1999).

Finally we use $C_2 = 0.2$ F m$^{-2}$ for quartz–water interface and for calcite–water interface, and $C_1 = 1.4$ F m$^{-2}$ or 0.81 F m$^{-2}$ for quartz–water interface, and $C_1 = 1.4$ F m$^{-2}$ for calcite–water interface. The variability of the integral capacitance values on the calculated zeta potentials is shown in Fig. 9(a).